\documentstyle{article}
\input amssym.def
\input amssym
\def\ol{\overline}
\def\tr{\mbox{Tr}}
\def\tp#1{#1^{\rm T}}
\def\dim{\mbox{dim}\;}
\def\a{\alpha}
\def\b{\beta}
\def\c{\gamma}
\def\d{\delta}
\def\s{\sigma}
\def\w{\omega}
\def\W{\Omega}
\def\fa{\frak A}
\def\fc{\frak C}
\def\dg{\dagger}
\def\ds{\displaystyle}
\def\mb#1{\mbox{\boldmath{$#1$}}}
\newtheorem{thm}{Theorem}
\newcommand{\be}{\begin{equation}}
\newcommand{\ee}{\end{equation}}
\newcommand{\bea}{\begin{eqnarray}}
\newcommand{\eea}{\end{eqnarray}}
\newcommand{\beax}{\begin{eqnarray*}}
\newcommand{\eeax}{\end{eqnarray*}}
\newcommand{\op}[1]{\mbox{\sf #1}}
\newcommand{\komm}[2]{\left[#1,#2\right]}
\newcommand{\akomm}[2]{\left\{#1,#2\right\}}

\newcommand{\bt}{\begin{thm}{\hspace{-.55em}\em{\bf {: }}}}
\newcommand{\et}{\end{thm}}

\begin{document}
\bibliographystyle{plain}
\begin{titlepage}
\renewcommand{\thefootnote}{\fnsymbol{footnote}}
\large
\hfill\begin{tabular}{l}HEPHY-PUB 675/97\\ UWThPh-1997-40\\ hep-th/9710202\\
October 1997\end{tabular}\\[3cm]
\begin{center}
{\Large\bf FINITE QUANTUM FIELD THEORIES:}\\[.5ex]
{\Large\bf CLIFFORD ALGEBRAS FOR YUKAWA COUPLINGS?}\\
\vspace{2.5cm}
{\Large\bf Wolfgang LUCHA\footnote[1]{\normalsize\ E-mail:
v2032dac@awiuni11.edvz.univie.ac.at}}\\[.5cm]
Institut f\"ur Hochenergiephysik,\\
\"Osterreichische Akademie der Wissenschaften,\\
Nikolsdorfergasse 18, A-1050 Wien, Austria\\[1.7cm]
{\Large\bf Michael MOSER\footnote[2]{\normalsize\ E-mail:
mmoser@galileo.thp.univie.ac.at}}\\[.5cm]
Institut f\"ur Theoretische Physik,\\
Universit\"at Wien,\\
Boltzmanngasse 5, A-1090 Wien, Austria\\[2cm]
{\bf Abstract}
\end{center}
\normalsize
By imposing on the most general renormalizable quantum field theory the
requirement of the absence of ultraviolet-divergent renormalizations of the
physical parameters (masses and coupling~constants) of the theory, finite
quantum field theories in four space-time dimensions may be constructed.
Famous ``prototypes'' of these form certain well-known classes of
supersymmetric finite quantum field theories. Within a perturbative
evaluation of the quantum field theories under consideration, the starting
point of all such investigations is represented by the conditions for one-
and two-loop finiteness of the gauge couplings as well as for one-loop
finiteness of the Yukawa couplings. Particularly attractive~solutions~of the
one-loop Yukawa finiteness condition involve Yukawa couplings which are
equivalent to generators of Clifford algebras with identity element. However,
our closer inspection shows, at least for all simple gauge groups up to and
including rank 8, that Clifford-like solutions prove to be inconsistent with
the requirements of one- and two-loop finiteness of the gauge coupling and of
absence of gauge anomalies.

\vspace*{2ex}

\normalsize
\renewcommand{\thefootnote}{\arabic{footnote}}
\end{titlepage}

\section{Introduction}

The standard theory of elementary particle physics, in spite of its enormous
success in describing the strong and electroweak interactions, exhibits a
very unpleasant feature, which it shares with~almost~all quantum field
theories: the appearance of ``ultraviolet divergences,'' order by order in
the perturbative loop expansion. Of course, within the subset of
renormalizable theories these divergences may be dealt with by application of
the so-called renormalization programme. Nevertheless, the ultimate goal here
should be an understanding of nature in terms of a {\em finite\/} theory,
i.e., a theory without any divergence.

Every additional symmetry is potentially able to improve the high-energy
behaviour of a quantum field theory---as may be seen by increasing gradually
the number $N$ of supersymmetries of the theory:
\begin{itemize}
\item All one-loop finite $N=1$ supersymmetric theories are (at least)
two-loop finite \cite{N=1}, even if this $N=1$ supersymmetry is softly broken
(in a well-defined way) \cite{N=1-soft}. Under certain circumstances, $N=1$
supersymmetric theories may be finite to all orders of their perturbative
expansion \cite{piguet}.
\item All $N=2$ supersymmetric theories satisfying merely one single
``finiteness condition'' are finite to all orders of the perturbative
expansion \cite{N=2}, even if one or both supersymmetries are softly broken
(in a well-defined way) \cite{N=2-soft}; these theories have been classified
under various aspects \cite{N=2-class}.
\item In the case of the $N=4$ supersymmetric Yang--Mills theory, that
``$N=2$ finiteness condition'' is trivially fulfilled by the particle content
of this theory enforced by $N=4$ supersymmetry \cite{N=4}.
\end{itemize}
The next target must be non-supersymmetric finite quantum field theories
\cite{lucha86-1,lucha86-2}: Is supersymmetry a necessary prerequisite for
finiteness? Do there exist non-supersymmetric finite quantum field theories?
A fundamental result specifying the particle content of finite quantum field
theories in four \mbox{space-time} dimensions has immediately been found
\cite{lucha86-1,lucha86-2,boehm87,lucha87-1}: {\em Any non-trivial finite
quantum field theory~must necessarily comprise vector bosons related to a
non-Abelian gauge group, fermions, {\em and\/} scalar bosons.} However,
the analysis of specific (classes of) models revealed, for instance, that
models being finite in dimensional regularization, at least up to some loop
order, may be plagued by quadratic divergences~in cut-off regularization
\cite{qdiv1,qdiv2}.

In all searches for non-supersymmetric finite quantum field theories, the
first genuine hurdle~to~be taken is the condition for one-loop finiteness of
the Yukawa couplings necessarily present in the~theory. A particular class of
solutions of this one-loop Yukawa finiteness condition is characterized by
Yukawa couplings which are equivalent to the generators of some Clifford
algebra $\fc$ with identity element~\cite{kranner90}. It has been speculated
\cite{kranner91} that finite theories involving these Clifford-like Yukawa
couplings might~be constructed. The intention of the present analysis is to
scrutinize systematically the relevance of these Clifford-like Yukawa
solutions for the construction of new, i.e., non-supersymmetric, finite
quantum field theories on a rather general basis. Details of this
investigation may be found in Refs.~\cite{lucha96-0,lucha96-1,lucha96-2}.
The C package developed in order to perform the numerical scan through
possible candidates for~finite theories is extensively described in
Ref.~\cite{package}.

\section{Finiteness Conditions in General Quantum Field
Theories}\label{sec:fcigqft}

Let us start from the most general \cite{llewellyn73} renormalizable quantum
field theory (for particles up to spin 1 $\hbar$) invariant with respect to
gauge transformations forming some compact simple Lie group $G$ with
corresponding Lie algebra $\fa$. The particle content of this theory is
described by
\begin{itemize}
\item (gauge) vector-boson fields $A_\mu(x)=(A_\mu^a)(x)\in\fa$ in the
adjoint representation $R_{{\rm ad}}$ of the gauge group $G$, of dimension
$d_{\rm g}:=\dim\fa$;
\item two-component (Weyl) fermion fields $\psi(x)=(\psi^i)(x)\in V_{\rm F}$
in some arbitrary representation~$R_{\rm F}$ of $G$, of dimension $d_{\rm
F}:=\dim V_{\rm F}$; and
\item Hermitean scalar boson fields $\phi(x)=(\phi^\a)(x)\in V_{\rm B}$ in
some arbitrary {\em real\/} representation $R_{\rm B}$~of $G$, of dimension
$d_{\rm B}:=\dim V_{\rm B}$.
\end{itemize}
The Lagrangian defining this theory is given by
\beax
\cal L&=&-\:\frac{1}{4}\,F_{\mu\nu}^a\,F^{\mu\nu}_a
+i\,\ol{\psi}_i\,\ol{\s}^\mu\left[(D_\mu)_{\rm F}\,\psi\right]^i
+\frac{1}{2}\left[(D_\mu)_{\rm B}\phi\right]^\a
\left[(D^\mu)_{\rm B}\phi\right]_\a\\[1ex]
&&-\:\frac{1}{2}\,\phi^\a\,Y_{\a ij}\,\psi^i\,\psi^j
-\frac{1}{2}\,\phi_\a\,Y^{\dg\a ij}\,\ol{\psi}_i\,\ol{\psi}_j
-\frac{1}{4!}\,V_{\a\b\c\d}\,\phi^\a\,\phi^\b\,\phi^\c\,\phi^\d\\[1ex]
&&+\:\mbox{mass terms}+\mbox{cubic scalar-boson self-interactions}
+\mbox{gauge-fixing and ghost terms}\ .
\eeax
The $d_{\rm g}$ Hermitean generators $T_R^a$, $a=1,2,\dots,d_{\rm g}$, of the
Lie group $G$ in an arbitrary, maybe reducible representation $R$ satisfy the
commutation relations
$$
\komm{T_R^a}{T_R^b}=i\,{f^{ab}}_c\,T_R^c\ ,
$$
with the structure constants ${f^{ab}}_c$, $a,b,c=1,2,\dots,d_{\rm g}$,
defining the Lie algebra $\fa$ under consideration. The gauge coupling
constant is denoted by $g$. The gauge-covariant field strength tensor
$F_{\mu\nu}^a$ is of the usual form,
$$
F_{\mu\nu}^a=\partial_\mu A_\nu^a-\partial_\nu A_\mu^a
+g\,{f^a}_{bc}\,A_\mu^b\,A_\nu^c\ .
$$
The gauge-covariant derivatives $D_\mu$ acting on the representation spaces
$V_{\rm F}$ and $V_{\rm B}$, respectively, read
$$
(D_\mu)_{\rm R}:=\partial_\mu-i\,g\,T_{\rm R}^a\,A_\mu^a\quad\mbox{for}\quad
{\rm R}=\mbox{F, B}\ .
$$
The four $2\times 2$ matrices $\ol{\s}^\mu$ in the kinetic term for the Weyl
fermion fields are defined in terms of the $2\times 2$ unit matrix,
$\op{1}_2$, and the three Pauli matrices, $\mb{\s}$, by
$\ol{\s}^\mu=(\op{1}_2,-\mb{\s})$. Without loss of generality, all the Yukawa
couplings $Y_{\a ij}$ may be assumed to be completely symmetric in their
fermionic indices $i$ and $j$, and all the quartic scalar-boson
self-couplings $V_{\a\b\c\d}$ may be taken to be completely symmetric under
an arbitrary permutation of their indices. The group invariants for an
arbitrary representation $R$ of $G$ are defined in terms of the generators
$T_R^a$ of ${\fa}$ in this representation $R$ as usual: the quadratic Casimir
operator $C_R$ is given by
$$
C_R:=\sum_{a=1}^{d_{\rm g}}T_R^a\,T_R^a\ ,
$$
and the second-order Dynkin index $S_R$ is obtained from
$$
S_R\,\d^{ab}:=\tr\left(T_R^a\,T_R^b\right)\ .
$$
In the adjoint representation $R_{{\rm ad}}$, the Casimir eigenvalue $c_{\rm
g}$ equals the Dynkin index $S_{\rm g}$, i.e., $c_{\rm g}=S_{\rm g}$.

According to our understanding of ``finiteness'' of a general renormalizable
quantum field theory,\footnote{\normalsize\ For a more detailed discussion of
our notion of ``finiteness'' of arbitrary renormalizable quantum field
theories, consult, for instance, Refs.~\cite{lucha86-1,lucha86-2,lucha96-0}.}
finiteness is tantamount to the vanishing of the beta functions of all
physical parameters of this theory, in case of perturbative evaluation of
this quantum field theory order by order in its loop expansion. By
application of the standard renormalization procedure with the help of
dimensional regularization in the minimal-subtraction scheme, the relevant
finiteness conditions may be easily extracted \cite{cheng74,machacek}, see
also Refs.~\cite{lucha86-1,lucha86-2}:
\begin{itemize}
\item The condition for one-loop finiteness of the gauge coupling constant
$g$ reads
\be
22\,c_{\rm g}-4\,S_{\rm F}-S_{\rm B}=0\ .
\label{fin8}
\ee
\item Adopting this result, the condition for two-loop finiteness of the
gauge coupling constant~$g$~reads
\be
\tr_{\rm F}\left(C_{\rm F}\sum_{\b=1}^{d_{\rm B}}Y^{\dg\b}\,Y_\b\right)
-2\,g^2\left[\tr(C_{\rm F})^2+\tr(C_{\rm B})^2+d_{\rm g}\,c_{\rm g}\,(S_{\rm
F}-2\,c_{\rm g})\right]=0\ ,
\label{fin9}
\ee
where by $\tr_{\rm F}$ we indicate the partial trace over the fermionic
indices only.
\item The condition for one-loop finiteness of the Yukawa couplings $Y_{\a
ij}$ reads
\bea
&&\sum_{\b=1}^{d_{\rm B}}\left\{4\,Y_\b\,Y^{\dg\a}\,Y_\b+Y_\a\,Y^{\dg\b}\,Y_\b
+Y_\b\,Y^{\dg\b}\,Y_\a+Y_\b\,\tr_{\rm F}\left(Y^{\dg\a}\,Y_\b
+Y^{\dg\b}\,Y_\a\right)\right\}\nonumber\\[1ex]
&&-\;6\,g^2\left[Y_\a\,C_{\rm F}+\tp{\left(C_{\rm F}\right)}\,Y_\a\right]=0\ ;
\label{fin10}
\eea
we call this (cubic and thus troublesome) relation, for brevity, the ``Yukawa
finiteness condition'' (YFC).
\end{itemize}
These three (lowest-order) finiteness conditions for the gauge and Yukawa
couplings, Eqs.~(\ref{fin8}), (\ref{fin9}), and (\ref{fin10}), have been
identified as the central part of the whole set of finiteness conditions: any
investigation of (perturbative) finiteness of quantum field theories should
start from this set of equations \cite{kranner91}. (The very first term in
Eq.~(\ref{fin9}) constitutes the link between the two-loop gauge-coupling
finiteness condition (\ref{fin9}) and the relation one obtains when
multiplying the YFC (\ref{fin10}) by $Y^{\dg\a}$, performing the sum over all
$\a=1,\dots,d_{\rm B}$, and taking the trace of the resulting expression with
respect to the fermionic indices.)

\section{The Standard Form of the Yukawa Finiteness Condition}\label{sec:yfc}

The YFC (\ref{fin10}) is obviously invariant under arbitrary ${\rm U}(d_{\rm
F})\otimes{\rm O}(d_{\rm B})$ transformations \cite{kranner91}. Luckily,~this
invariance and the gauge invariance of the Yukawa couplings $Y_{\a ij}$
enforced by the gauge invariance~of our Lagrangian conspire to render
possible the {\em simultaneous diagonalization\/} of the Casimir operators
$C_{\rm F}$ and $C_{\rm B}$, which results in
\beax
{(C_{\rm F})^i}_j&=&{\d^i}_j\,C_{\rm F}^j\ ,\\[1ex]
{(C_{\rm B})^\a}_\b&=&{\d^\a}_\b\,C_{\rm B}^\b\ ,
\eeax
and certain sesquilinear products of the Yukawa couplings $Y$, namely, the
bosonic and fermionic~traces
\beax
\sum_{\b=1}^{d_{\rm B}}{\left(Y^{\dg\b}\,Y_{\b}\right)^i}_j
&=&{\d^i}_j\,y_{\rm F}^j\ ,\\[1ex]
\tr_{\rm F}\left(Y^{\dg\a}\,Y_{\b}+Y^{\dg\b}\,Y_{\a}\right)
&=&2\,{\d^\a}_\b\,y_{\rm B}^\b\ .
\eeax
With all these eigenvalues, the YFC (\ref{fin10}) simplifies to what may be
regarded its {\em standard form\/} \cite{kranner91,lucha96-0}:
\be
4\sum_{\b=1}^{d_{\rm B}}\left(Y_\b\,Y^{\dg\a}\,Y_\b\right)_{ij}
+Y_{\a ij}\left(2\,y_{\rm B}^\a+y_{\rm F}^i+y_{\rm F}^j-6\,g^2\,C_{\rm F}^i
-6\,g^2\,C_{\rm F}^j\right)=0\ .
\label{st1}
\ee

In order to explore the implications of gauge invariance for the Yukawa
couplings, we decompose the bosonic index $\a$ and the fermionic index $i$
into pairs of indices, $\a=(A,\a_A)$ and $i=(I,i_I)$, where the indices $A$
and $I$ distinguish the irreducible representations $R_{\rm B}^A\subset
R_{\rm B}$ and $R_{\rm F}^I\subset R_{\rm F}$, respectively, while the
indices $\a_A=1,\dots,d_A$ and $i_I=1,\dots,d_I$ label the components of
$R_{\rm B}^A$ and $R_{\rm F}^I$, respectively. If and only if the product
$R_{\rm B}^A\otimes R_{\rm F}^I\otimes R_{\rm F}^J$ of any three irreducible
representations $R_{\rm B}^A\subset R_{\rm B}$, $R_{\rm F}^I\subset R_{\rm
F}$, and $R_{\rm F}^J\subset R_{\rm F}$ of $G$ contains the trivial
representation, $\op{1}$, $N^{(A,I,J)}$ times, there exist $N^{(A,I,J)}$
invariant tensors $(\Lambda^{(k)})_{\a_A i_I j_J}$. In terms of these
tensors, the gauge-covariant expansion of $Y$, with coefficients
$p^{(k)}_{AIJ}\in\Bbb C$, reads
\be
Y_{\a ij}=Y_{(A,\a_A)(I,i_I)(J,j_J)}=\sum_{k=1}^{N^{(A,I,J)}}p^{(k)}_{AIJ}
\left(\Lambda^{(k)}\right)_{\a_Ai_Ij_J}\ .
\label{red7}
\ee

Following Ref.~\cite{kranner91}, we introduce a certain---and, upon
application of the two-loop gauge-coupling finiteness condition (\ref{fin9}),
purely group-theoretic---quantity $F^2$, by defining
\be
F^2:=\frac{\ds\tr_{\rm F}\left(C_{\rm F}\ds\sum_{\b=1}^{d_{\rm
B}}Y^{\dg\b}\,Y_\b\right)}{\ds 6\,g^2\,\tr(C_{\rm F})^2}
=\frac{\ds\tr(C_{\rm F})^2+\tr(C_{\rm B})^2+d_{\rm g}\,c_{\rm g}\,(S_{\rm F}
-2\,c_{\rm g})}{\ds 3\,\tr(C_{\rm F})^2}\ .
\label{fin29}
\ee
Remarkably, each theory which satisfies the central part of finiteness
conditions, Eqs.~(\ref{fin8}), (\ref{fin9}), and~(\ref{fin10}), also
satisfies the inequality $F^2\leq 1$. In particular, the extremum $F^2=1$
seems to play a decisive r\^ole in the analysis of these finiteness
conditions \cite{kranner91}: For $F^2=1$ and only for this case, the cubic
YFC (\ref{fin10}) simplifies to a merely quadratic system, which is fulfilled
by every $N=1$ supersymmetric~(two-loop-) finite theory. Numerical
investigations \cite{kranner91} revealed that for finite quantum field
theories the~value~of $F$ is close to $F^2=1$. These findings led to
conjecture \cite{kranner91} that all finite theories might satisfy $F^2=1$.

We call a quantum field theory ``potentially finite'' if its particle content
fulfills both the finiteness condition (\ref{fin8}) and the inequalities
$0<F^2\leq 1$ for the quantity $F^2$ as defined by Eq.~(\ref{fin29}), if the
anomaly index of its fermionic representation, $R_{\rm F}$, vanishes, if its
bosonic representation, $R_{\rm B}$, is real, $R_{\rm B}\simeq R_{\rm B}^*$,
and if, at least, one fundamental invariant tensor, required for the
decomposition (\ref{red7}) of $Y_{\a ij}$, exists.

\section{($\W$-Fold) Reducibility of the Yukawa Finiteness
Condition}\label{sec:reduc}

We re-order both the fermionic indices $i$ and the bosonic indices $\a$ such
that the first $n\leq d_{\rm F}$ fermionic indices and the first $m\leq
d_{\rm B}$ bosonic indices cover precisely those subsets of $R_{\rm F}$ and
$R_{\rm B}$, respectively, which have non-vanishing Yukawa couplings. This
ordering is then equivalent to the requirement \cite{lucha96-0}
\beax
y_{\rm F}^i\neq 0&\Leftrightarrow&i\in\{1,2,\dots,n\leq d_{\rm F}\}\ ,\\[1ex]
y_{\rm B}^\a\neq 0&\Leftrightarrow&\a\in\{1,2,\dots,m\leq d_{\rm B}\}\ .
\eeax

Let us call any two sets $M_q=\{(R^{\mu_q},R^{I_q},R^{J_q})\}$, $q=1,2,$ of
combinations of real bosonic blocks\footnote{\normalsize\ Since the bosonic
representation $R_{\rm B}$ must be real, every non-orthogonal irreducible
representation $R_{\rm B}^A\subset R_{\rm B}$ has to find a mutually
contragredient companion $(R_{\rm B}^A)^{\rm c}\subset R_{\rm B}$ in order to
be able to form~a real orthogonal block: $R_{\rm B}^\mu\simeq R_{\rm
B}^A\oplus (R_{\rm B}^A)^{\rm c}$.} $R^{\mu_q}\subset R_{\rm B}$ and
irreducible fermionic representations $R^{I_q},R^{J_q}\subset R_{\rm F}$
appearing in $Y_{(\mu,\a_\mu)(I,i_I)(J,j_J)}$ to be disjoint if and only if
$\{R^{\mu_1}\}\cap\{R^{\mu_2}\}=\{R^{I_1}\}\cap\{R^{I_2}\}=\{R^{J_1}\}\cap
\{R^{J_2}\}=\emptyset$. With this, we define \cite{lucha96-1}: Let
$M=\{(R^\mu,R^I,R^J)\mid R^\mu\subset R_{\rm B},\ R^I,R^J\subset R_{\rm F}\}$
be the set of all combinations of real bosonic blocks and irreducible
fermionic representations in the YFC (\ref{st1}). If $M$ is the union of
$\W\geq 1$ pairwise disjoint non-empty subsets $M_\w$, $\w=1,2,\dots,\W$,
that is,
$$
M=\bigcup_{\w=1}^\W M_\w\ ,
$$
we call this YFC {\em $\W$-fold reducible}. Rather trivially, an only {\em
1-fold\/} reducible YFC is called {\em irreducible}.\footnote{\normalsize\
Interestingly, for supersymmetric theories the YFC (\ref{st1}) is always
irreducible \cite{moser97}. An example for a reducible YFC has been
constructed by considering some {\em non-supersymmetric\/} particle
content~\cite{skarke94-2}.} $\W$-fold reducibility of the YFC splits both the
``fermionic'' representation space $V_{\rm F}$ and the ``bosonic''
representation space $V_{\rm B}$ into direct sums of subspaces, with
corresponding ``fermionic'' and ``bosonic'' dimensions $n_\w$ and $m_\w$,
respectively; each of these subspaces is related to some irreducible
component $M_\w$. For every of these irreducible components $M_\w$, the YFC
(\ref{st1}) assumes its reduced standard form:
\be
4\sum_{\b=1}^{m_\w}\left(Y_\b\,Y^{\dg\a}\,Y_\b\right)_{ij}
+Y_{\a ij}\left(2\,y_{\rm B}^\a+y_{\rm F}^i+y_{\rm F}^j-6\,g^2\,C_{\rm F}^i
-6\,g^2\,C_{\rm F}^j\right)=0
\label{red5}
\ee
for $1\leq i,j\leq n_\w$ and $1\leq\a\leq m_\w$.

The constraint $F^2=1$ may also be expressed by requiring $y_{\rm
F}^i=6\,g^2\,C_{\rm F}^i$ for all $i\in\{1,\dots,n=d_{\rm F}\}$. More
generally, we may demand $y_{\rm F}^i=6\,g^2\,C_{\rm F}^i$ for all
$i\in\{1,\dots,n<d_{\rm F}\}$ and, of course, $y_{\rm F}^i=0$ for~all
$i\in\{n+1,\dots,d_{\rm F}\}$. We then arrive at a situation similar to the
case $F^2=1$ but now with $F^2<1$:~the cubic YFC (\ref{st1}) simplifies also
in this case to a quadratic system. With the help of our C package
\cite{package}, the existence of potentially finite theories solving the
one-loop gauge-coupling finiteness condition~(\ref{fin8}) and the latter
quadratic system may be shown numerically \cite{moser97}. Hence, the
advantages brought about by the particular value $F^2=1$ are shared by models
with values of $F^2$ different from this special~case.

\section{Symmetries of the Yukawa Finiteness Condition}\label{sec:sym}

Since the term which causes all these troubles in the analysis of the YFC is
the first (cubic) expression on the left-hand side of Eq.~(\ref{st1}), we
focus our attention to the investigation of a quantity $x$ defined~by
$$
2\,{x^{i\a}}_{j\b}:={\left(Y^{\dg\a}\,Y_\b+Y^{\dg\b}\,Y_\a\right)^i}_j\ .
$$
Our basic building blocks are the irreducible components $M_\w$; for any
index $\a$ or $i$ corresponding~to~a given $M_\w$, we simply write $\a\in
M_\w$ and $i\in M_\w$, respectively. Every component $M_\w$ is invariant
under all ${\rm U}(n_\w)\otimes{\rm O}(m_\w)$ transformations. Let's assume
that the corresponding irreducible component $x_\w$~of $x$ is diagonalizable
by some transformation $S_\w$ of this kind (for a detailed discussion,
consult Ref.~\cite{lucha96-2}):
$$
{(x_\w)^{i\a}}_{j\b}
\stackrel{{\rm U}(n_\w)\otimes{\rm O}(m_\w)}{\longrightarrow}
{\left(S_\w\,x_\w\,S_\w^\dg\right)^{i\a}}_{j\b}=
{\d^i}_j\,{\d^\a}_\b\,x_\w^{j\b}\quad\forall\ i,j,\a,\b\in M_\w\ .
$$
This transformation enables us to cast the YFC (\ref{red5}) into a form
quasi-linear in the Yukawa couplings:
\be
Y_{\a ij}
\left(4\,x_\w^{i\a}+4\,x_\w^{j\a}+2\,y_{\rm B}^\a-y_{\rm F}^i-y_{\rm F}^j-
6\,g^2\,C_{\rm F}^i-6\,g^2\,C_{\rm F}^j\right)=0\quad\forall\ \a,i,j\in M_\w\ .
\label{cred5}
\ee

A particularly important subset of $S_\w$-diagonalizable quantities $x_\w$ is
given by $x_\w$ being the tensor product of a factor, $u_\w$, carrying only
bosonic indices and a factor, $v_\w$, carrying only fermionic indices:
$x_\w=u_\w\otimes v_\w$. For $x_\w$ tensorial, our situation simplifies
drastically: one easily proves the relation~\cite{lucha96-0}
$$
{(x_\w)^{i\a}}_{j\b}={\d^i}_j\,{\d^\a}_\b\,x_\w^{j\b}=
{\d^i}_j\,{\d^\a}_\b\,\frac{y_{\rm F}^j\,y_{\rm B}^\b}
{\sum_{k\in M_\w}y_{\rm F}^k}\quad\forall\ i,j,\a,\b\in M_\w\ ,
$$
from which one learns that every irreducible component $x_\w$ of $x$ is
invertible, and the commutator~\cite{lucha96-0}
$$
Y_{\a ij}\left(y_{\rm F}^i-y_{\rm F}^j\right)=0\quad\forall\ \a,i,j\in M_\w\ .
$$

\newpage
The solution of the quasilinear YFC for $x_\w^{i\a}$ then involves some sort
of average $C_{\rm m}^i=(C_{\rm F}^i+C_{\rm F}^{j(i)})/2$ of fermionic
Casimir eigenvalues (where, as indicated by our notation, the fermionic index
$j$ depends on $i$ but not on $\a$):
\be
\left(4-m_\w\right)x_\w^{i\a}=6\,g^2\left(C_{\rm m}^i-\frac{1}{4-m_\w+\,n_\w}
\sum_{k\in M_\w}C_{\rm m}^k\right)\quad\forall\ i,\a\in M_\w\ .
\label{sy4}
\ee
Consequently, for a tensorial $x_\w$, the corresponding eigenvalues are
independent of $\a\in M_\w$: $x_\w^{i\a}=x^i$.

Now, what about Clifford algebras in this context? It is rather
straightforward to prove \cite{kranner90,lucha96-0,moser97} that any set of
matrices $Y_\a$ satisfying the relations
$$
{\left(Y^{\dg\a}\,Y_\b+Y^{\dg\b}\,Y_\a\right)^i}_j=
2\,{\d^\a}_\b\,{\d^i}_j\,x^j\quad\forall\ \a,\b,i,j\in M_\w
$$
is equivalent to the union of the $n_\w\times n_\w$ unit matrix
$\op{1}_{n_\w}$ and the subset
$$
{\frak B}_\w=\{N_\a\mid\akomm{N_\a}{N_\b}=2\,\d_{\a\b}\,\op{1}_{n_\w},\
N_{\a ij}=N_{\a ji}\in{\Bbb R},\ \a=1,\dots,m_\w-1\}
$$
of real, symmetric, and anticommuting elements $N_\a$ of a representation of
some Clifford algebras $\fc_i$. This Clifford algebra structure restricts,
for any irreducible component $M_\w$, the possible ranges of~the respective
bosonic and fermionic dimensions $m_\w$ and $n_\w$: The rank $p_i$ of some
Clifford algebra~${\fc}_i$ may be either even, $p_i=2\,\nu_i$, or odd,
$p_i=2\,\nu_i+1$, with $\nu_i\in\Bbb N$. In both cases, any matrix
representation~of ${\fc}_i$ is built from $2^{\nu_i}$-dimensional blocks, and
there exist precisely $q_i=\nu_i+1$ symmetric anticommuting elements.
Demanding to have enough of these elements at one's disposal translates into
the inequality \cite{lucha96-0}
\be
n_\w\geq 2^{m_\w-2}\ .
\label{sy6}
\ee

\section{Representations of Clifford Algebras for $F^2=1$ Theories}

The restrictivity of Inequality (\ref{sy6}) may be demonstrated by applying
it to the class of $F^2=1$~theories:
\bt
Let the YFC be\/ $\W$-fold reducible and assume $x_\w=u_\w\otimes v_\w$,
$1\leq\w\leq\W$; then there does not exist any $F^2=1$ solution of the YFC
obeying the following criteria:
\begin{enumerate}
\item The fermionic representation $R_{\rm F}$ has vanishing anomaly index.
\item The bosonic representation $R_{\rm B}$ is real (orthogonal).
\item The beta function for the gauge coupling $g$ vanishes in one-loop
approximation.
\end{enumerate}
\et

In order to prove the above statement, we employ the subroutine {\tt
constraint} of our C package~\cite{package} to implement the requirements
$F^2=1$ and
$$
n=\sum_{\w=1}^{\W}n_\w=d_{\rm F}\ .
$$
The bosonic and fermionic dimensions of every irreducible component $M_\w$ of
the YFC are related by
\be
4+n_\w=2\,m_\w\ ,
\label{f1}
\ee
indicating that any fermionic dimension $n_\w$ must be even. This relation
may then be used to eliminate the bosonic dimension $m_\w$ from the
inequality (\ref{sy6}), with the result
$$
n_\w\geq 2^{m_\w-2}=2^{n_\w/2}\ ,
$$
from which we deduce that the fermionic dimension $n_\w$ is necessarily
restricted to one of three values: $n_\w=2,3,4$. Consequently, for
potentially finite $F^2=1$ theories with Clifford-type Yukawa couplings,
there are no more than two options for the dimensions of any irreducible
component $M_\w$ of the YFC: $(n_\w=2,m_\w=3)$ or $(n_\w=4,m_\w=4)$. Needless
to say, every irreducible component $M_\w$ of the YFC has to embrace both
complete irreducible fermionic representations $R_{\rm F}^I$ and complete
real orthogonal bosonic blocks $R_{\rm B}^\mu$ of representations of the Lie
algebra $\fa$, coupling invariantly within the component $M_\w$. Direct
inspection of all simple Lie algebras $\fa$ shows that only the four Lie
algebras~$A_1$,~$A_2$,~$A_3$, and $B_2$ possess irreducible representations
of sufficiently low dimension for use in the fermionic sector. The following
case-by-case examination of all potentially finite theories extracted in this
manner then allows us to claim that there are no potentially finite $F^2=1$
solutions of the quasi-linear irreducible YFC (\ref{cred5}) obeying
simultaneously Inequality (\ref{sy6}) for corresponding bosonic and fermionic
dimensions. (For the purpose of these analyses, let us denote any
$d$-dimensional irreducible representation by $[d]$.)

\subsection{The Lie Algebra A$_1$}

For the Lie algebra A$_1$, the only irreducible representations of dimensions
less than or equal to 4 are the two-, three-, and four-dimensional
representions [2], [3], [4]. All potentially finite $F^2=1$ theories based on
A$_1$ with fermionic representations $R_{\rm F}$ containing only these three
irreducible representations are listed (consecutively numbered) in
Table~\ref{tab:a1}. The appearance of (any number of) three-dimensional
irreducible representations in the fermionic representation $R_{\rm F}$ of a
potentially finite theory is certainly incompatible with either of the two
conceivable values, $n_\w=2$ or $n_\w=4$, of the fermionic dimension $n_\w$
of any irreducible component $M_\w$ of our YFC. Inspection of
Table~\ref{tab:a1} leaves us with two candidates:
\begin{itemize}
\item Theory no.~1 is consistent with the requirement $d_{\rm F}\leq 4$,
valid for an irreducible YFC. Invariant tensors to construct gauge-invariant
Yukawa couplings exist only for $[3]\otimes[4]\otimes[4]$. Therefore, $m$ may
take values in $\{3,6,9,\dots,21\}$, whereas Eq.~(\ref{f1}) for $d_{\rm F}=4$
implies $m=4$.
\item Theory no.~4 involves the four-dimensional irreducible fermion
representation [4], to be covered by an irreducible component $M_\w$ of
fermionic dimension $n_\w=4$, which, in turn, implies $m_\w=4$ for its
bosonic dimension. However, since $[4]\otimes[4]\not\supset[2]$, there is no
suitable invariant tensor~$\Lambda^{(k)}$.
\end{itemize}
Thus there remains no candidate for a Clifford-type finite $F^2=1$ theory
based on the Lie algebra~A$_1$.

\begin{table}[htb]
\caption{Potentially finite $F^2=1$ theories for the Lie algebra A$_1$ with
fermionic representations $R_{\rm F}$ involving only irreducible
representations of dimension less than or equal to 4. The multiplicities of a
$d$-dimensional irreducible representation of A$_1$ in $R_{\rm F}$ and
$R_{\rm B}$ are denoted by $f_{[d]}$ and $b_{[d]}$,
respectively.}\label{tab:a1}
\begin{center}
\begin{tabular}{rrrrrrr}
\hline\hline\\[-1.5ex]
Theory no.&$\qquad f_{[2]}$&$f_{[3]}$&$f_{[4]}$&$\qquad
b_{[2]}$&$b_{[3]}$&$b_{[4]}$\\[1ex]
\hline\\[-1.5ex]
1&0&0&1&20&7&0\\
2&0&3&0&32&2&0\\
3&0&4&0&0&6&0\\
4&2&0&1&8&8&0\\
5&2&1&1&4&0&2\\
6&2&3&0&20&3&0\\
7&4&2&0&40&0&0\\
8&4&3&0&8&4&0\\
9&6&2&0&28&1&0\\
10&8&2&0&16&2&0\\
11&10&2&0&4&3&0\\
12&12&1&0&24&0&0\\
13&14&1&0&12&1&0\\
14&16&1&0&0&2&0\\[1ex]
\hline\hline
\end{tabular}
\end{center}
\end{table}

\subsection{The Lie Algebra A$_2$}

For the Lie algebra A$_2$, the only irreducible representation of dimension
less than or equal to 4 is the three-dimensional fundamental representation
[3]. Obviously, it is not possible to construct invariant $n_\w=2$ or
$n_\w=4$ blocks from three-dimensional representations only. This fact rules
out any theory based on A$_2$.

\subsection{The Lie Algebra A$_3$}

For the Lie algebra A$_3$, the only irreducible representation of dimension
less than or equal to 4 is the (four-dimensional) fundamental representation
[4]. However, there exists no potentially finite $F^2=1$ theory with a
fermionic representation $R_{\rm F}$ which involves only this representation
[4]; in other words, every fermionic representation $R_{\rm F}$ in
potentially finite $F^2=1$ theories based on A$_3$ contains at least one
irreducible representation of dimension greater than 4. This circumstance
rules out every theory based on A$_3$.

\subsection{The Lie Algebra B$_2$}

For the Lie algebra B$_2$, the only irreducible representation of dimension
less than or equal to 4 is the four-dimensional fundamental representation,
[4]. All potentially finite $F^2=1$ theories based on B$_2$ with fermionic
representations $R_{\rm F}$ involving only this irreducible representation
are listed in Table~\ref{tab:b2}; all corresponding bosonic representations
$R_{\rm B}$ involve the four-, five-, and ten-dimensional irreducible
representations [4], [5], [10] of B$_2$. Every four-dimensional irreducible
representation in the fermionic representation of any of these candidate
theories must be covered by an irreducible component $M_\w$ of fermionic
dimension $n_\w=4$. This, in turn, fixes the bosonic dimension of this
particular irreducible component $M_\w$ to the value $m_\w=4$. However, the
tensor product of two four-dimensional irreducible representations [4] does
not contain the four-dimensional irreducible representation [4]:
$[4]\otimes[4]\not\supset[4]$. Consequently, no appropriate gauge-invariant
tensors may be constructed. We conclude that the Lie algebra B$_2$ provides
no viable candidate for a finite $F^2=1$ theory with Clifford-like Yukawa
couplings.

\begin{table}[htb]
\caption{Potentially finite $F^2=1$ theories for the Lie algebra B$_2$ with
fermionic representations $R_{\rm F}$ involving only irreducible
representations of dimension less than or equal to 4. The multiplicities of a
$d$-dimensional irreducible representation of B$_2$ in $R_{\rm F}$ and
$R_{\rm B}$ are denoted by $f_{[d]}$ and $b_{[d]}$,
respectively.}\label{tab:b2}
\begin{center}
\begin{tabular}{rrrrr}
\hline\hline\\[-1.5ex]
Theory no.&$\qquad f_{[4]}$&$\qquad b_{[4]}$&$b_{[5]}$&$b_{[10]}$\\[1ex]
\hline\\[-1.5ex]
1&29&14&1&0\\
2&30&4&4&0\\
3&31&2&0&1\\[1ex]
\hline\hline
\end{tabular}
\end{center}
\end{table}

\section{Clifford-Algebra Representations for Arbitrary Theories}

In principle, it is straightforward to solve the YFC in the form
(\ref{cred5}) for arbitrary values of $F^2$ \cite{lucha96-0,lucha96-1}. The
only quantity in Eq.~(\ref{cred5}) which does not depend on the Yukawa
couplings $Y_{\a ij}$ is the expression $6\,g^2\,C_{\rm F}$, which is also
independent of $\a$. Furthermore, because of the (highly welcome)
quasi-linearity of the YFC (\ref{cred5}), for this set of equations to be
solvable at all, the quantities $x_\w^{i\a}$ must be of the order
$\mbox{O}(g^2)$; that is, all components $x_\w^{i\a}$ of $x$, if regarded as
functions of $6\,g^2\,C_{\rm F}^i$, have to be quadratic~in~the gauge
coupling constant $g$. Accordingly, we adopt the---with these insights rather
reasonable---{\em ansatz\/}
$$
x_\w^{i\a}=x^i=6\,g^2\,a_\w\,C_{\rm F}^i+b_\w\quad\mbox{for}\quad
a_\w,b_\w\in{\Bbb C}\ ,\ 1\leq\w\leq\W\ ,\ i,\a\in M_\w\ .
$$
Then, for an $\W$-fold reducible YFC with tensorial $x_\w$ for all
irreducible components $M_\w$, $\w=1,\dots,\W$, the solutions for any given
component $M_\w$ may be classified according to the following
characteristics:
\begin{enumerate}
\item[A:] For $a_\w=0$, only one common value for all $y_{\rm F}^i$,
proportional to some average $C_{\rm m}^\w:=(C_{\rm F}^i+C_{\rm F}^j)/2$ of
Casimir eigenvalues, is conceivable:
$$
y_{\rm F}^i\equiv y_\w=6\,g^2\,\frac{m_\w}{4-m_\w+n_\w}\,C_{\rm m}^\w\quad
\forall\ i\in M_\w\ .
$$
\item[B:] For $a_\w\neq(4-m_\w)^{-1}$, only one fermionic Casimir eigenvalue
$C_\w$ is admissible, ${(C_{\rm F})^i}_j={\d^i}_j\,C_\w$, and only one common
value $y_\w$ for all $y_{\rm F}^i$ is allowed:
$$
y_{\rm F}^i\equiv y_\w=6\,g^2\,\frac{m_\w}{4-m_\w+n_\w}\,C_\w\quad\forall\
i\in M_\w\ .
$$
\item[C:] For $a_\w=(4-m_\w)^{-1}$, different values for $y_{\rm F}^i$ are
possible:
$$
(4-m_\w)\,y_{\rm F}^i=6\,g^2\,m_\w\left(C_{\rm F}^i-\frac{1}{4-m_\w+n_\w}
\sum_{k\in M_\w}C_{\rm F}^k\right)\quad\forall i\in M_\w\ .
$$
\end{enumerate}
Of course, every Clifford-like solution of the YFC deduced in this way has to
be subjected to, at least, the additional requirements of one- and two-loop
finiteness of the gauge coupling $g$ as well as anomaly freedom of the
theory, in order to be considered a serious candidate for a
(non-supersymmetric) finite quantum field theory. Clearly, all the above
sets of solutions are consistent with the general result~(\ref{sy4}).

\newpage
In order to list all candidate theories of interest for us, we employ again
our C package \cite{package},~which provides us with all potentially finite
theories for any given simple Lie algebra $\fa$. We confine ourselves to
theories where all irreducible representations able to evolve invariant
tensors for Yukawa couplings, together with their respective partners, if
necessary, indeed contribute. The systematic search
\cite{lucha96-0,lucha96-1} for finite quantum field theories with
Clifford-like Yukawa couplings may be carried out
numerically.\footnote{\normalsize\ In the case of the Lie algebra $A_3$, that
is, for the Lie group ${\rm SU}(4)$, with representations involving the
antisymmetric irreducible representation, the required computer resources
become rather large.} We investigated all simple Lie algebras up to and
including rank 8. Our findings may be summarized by some sort of no-go
theorem \cite{lucha96-0,lucha96-1}:
\bt
Consider a simple Lie algebra\/
${\fa}\in\{\mbox{A}_r,\mbox{B}_r,\mbox{C}_r,\mbox{D}_r,\mbox{E}_6,\mbox{E}_7,
\mbox{E}_8,\mbox{F}_4,\mbox{G}_2\mid r=\mbox{rank}\,{\fa}\leq 8\}.$ Let the
YFC be\/ $\W$-fold reducible and let every irreducible component $M_\w$ of
the YFC be tensorial,~i.e., assume that, for $1\leq\w\leq\W$, the irreducible
components $x_\w$ of $x$ are of the tensor form $x_\w=u_\w\otimes v_\w$, with
$u_\w$ carrying only bosonic indices and $v_\w$ carrying only fermionic
indices. Let
$$
x_\w^{i\a}=x^i=6\,g^2\,a_\w\,C_{\rm F}^i+b_\w\quad\mbox{for}\quad
a_\w,b_\w\in{\Bbb C}\ ,\ 1\leq\w\leq\W\ ,\ i,\a\in M_\w\ .
$$
Then, for arbitrary $a_\w$, there does not exist any solution if the YFC is
irreducible (i.e., for\/ $\W=1$)~and, for $a_\w\neq 0$, there does not exist
any solution if the YFC is arbitrarily\/ $\W$-fold reducible, provided these
solutions are subject to the following requirements:
\begin{enumerate}
\item The fermionic representation $R_{\rm F}$ has vanishing anomaly index.
\item The bosonic representation $R_{\rm B}$ is real (orthogonal).
\item The beta function for the gauge coupling $g$ vanishes in one- and
two-loop approximation.
\item Irreducible blocks $R_{\rm B}^\mu\subset R_{\rm B}$ and $R_{\rm
F}^I,R_{\rm F}^J\subset R_{\rm F}$, with multiplicities $b_\mu$, $f_I$, and
$f_J$, respectively, which allow for invariant couplings, i.e., $R_{\rm
B}^\mu\otimes R_{\rm F}^I\otimes R_{\rm F}^J\supset\op{1}$, contribute to the
YFC such that
$$
\left.y_{\rm F}\right|_{\op{1}_{f_I}\times R^I}\neq 0
$$
and
$$
\left.y_{\rm B}\right|_{\op{1}_{b_\mu}\times R^\mu}\neq 0\ .
$$
\end{enumerate}
\et

\section{Summary and Conclusions}\label{sec:sc}

Motivated by the (recent) conjecture \cite{kranner90,kranner91} that some
particular set of solutions of the condition~for one-loop finiteness of the
Yukawa couplings in general renormalizable quantum field theories which is
characterized by the fact that the resulting Yukawa coupling matrices are
equivalent to the generators of a Clifford algebra with identity element
might allow to construct new finite quantum field theories, we checked the
possibility to find among all these models with Clifford-like Yukawa
couplings theories which solve, in addition, the conditions for one- and
two-loop finiteness of the gauge coupling constant as well as for absence of
gauge anomalies. Very surprisingly, for all gauge groups with rank less than
or equal to 8, we did not succeed to find any candidate for a finite quantum
field theory with Clifford-like Yukawa couplings (under few reasonable
assumptions about the structure of these Yukawa solutions). A Clifford
structure of the Yukawa couplings, despite perhaps ideally suited for solving
the one-loop Yukawa finiteness condition, appears to be incompatible already
with finiteness of the gauge coupling.

\section*{Acknowledgements}

M.~M.~was supported by the ``Fonds zur F\"orderung der wissenschaftlichen
Forschung in \"Osterreich,'' project 09872-PHY, by the Institute for High
Energy Physics of the Austrian Academy of Sciences, and by a grant of the
University of Vienna.

\end{document}